\documentclass[12pt]{article}
\usepackage{graphicx}
\usepackage{amsmath}%
\begin{document}
%%%%%%%%%%%%%%%%%%%%%%%%%%%%%%%%%%%%%%%%%%%%%%%%%%%%%%%%%%%%%%%%%%%%%%

%\begin{titlepage}
\footnotetext[1]{jinght@ihep.ac.cn}
\footnotetext[2]{shenpn@ihep.ac.cn}
\footnotetext[3]{chiang@ihep.ac.cn}

\title{Hypernucleus Production by $A(p,pK^+)_{\Lambda}B$ Reactions}
\date{}

\maketitle
\begin{center}
\author{Hantao Jing$^{1,6}$, Pengnian Shen$^{5,1,3,4}$, Huanching Chiang$^{1,2}$\\

{\small $^1$Institute of High Energy Physics, Chinese Academy of Sciences,}\\
{\small P.O.Box 918(4), Beijing 100049, P.R.China}\\
{\small $^2$South-west University, Chongqing 400715, China}\\
{\small $^3$Department of Physics, Guangxi Normal University, Guilin
541004, China}\\
{\small $^4$Institute of Theoretical Physics, Chinese Academy of
Sciences, P.O.Box 2735, P.R.China}\\
{\small $^5$Center of Theoretical Nuclear Physics, National
Laboratory of Heavy Ion Accelerator,}\\
{\small Lanzhou 730000, P.R.China}\\
{\small $^6$Graduate School of the Chinese Academy of Sciences,
Beijing 100049, P.R.China} }

\end{center}

\begin{abstract}
The $\Lambda$-hypernucleus production by $A(p, pK^+)_{\Lambda}B$
reactions is investigated within the framework of the distorted wave
impulse approximation(DWIA). The amplitude for the elementary
process is evaluated in a fully covariant two-nucleon model based on
the effective Lagrangian. The reaction cross sections for
$\Lambda$-hypernucleus productions on $^6Li$, $^{12}C$ and $^{16}O$
targets are calculated. It is found that the distortion effects tend
to reduce the cross sections by a factor of 3$\sim$10. Various
differential cross sections (DCS) and double differential cross
sections (DDCS) are presented. It is shown that for the
$s_{\Lambda}-$wave hypernucleus production, the DCS is decreased
with increasing nuclear mass, and the DCS for the $p_{\Lambda}-$wave
hypernucleus production is normally higher than that for the
$s_{\Lambda}-$wave hypernucleus production. As a reference, the DDCS
with respect to the momenta of the outgoing proton and kaon is also
demonstrated. Finally, the missing mass spectra of the inclusive
reaction $p+A\rightarrow p+ K^+  + X$ for $^6Li$, $^{12}C$ and
$^{16}O$ targets are presented, from which the masses of hypernuclei
can accurately be extracted. Thus, we conclude that the missing mass
spectrum method is an alternative to study hypernuclear physics. And
the study of hypernuclear physics can be carried out in COSY and CSR
by the $A(p,pK^+)_{\Lambda}B$ reaction due to the $\mu$b-order
reaction cross sections.

\vspace{1cm}

\noindent PACS: 25.40.-h, 21.80.+a, 13.75.-n

\vspace{1cm}

\noindent {\it Keywords}: Elementary process, hypernucleus
production, proton nucleus collisions
\end{abstract}
%\end{titlepage}
%%%%%%%%%%%%%%%%%%%%%%%%%%%%%%%%%%%%%%%%%%%%%%%%%%%%%%%%%%%%%%%%%%%%%%%%%%%

\vspace{1cm}
%section 1
\section{Introduction}
There have been extensive advances in the field of hypernuclear
physics. The conventional $(K^-,\pi ^-)$ strangeness exchange
reactions and the associated production reactions of hypernuclei
induced by energetic pion, photon and anti-proton beams have been
studied experimentally and theoretically  in last 30
years\cite{PLB83_306, PLB84_393, NPA360_315, PRC22_2073, PRC53_1210,
EPLA18_283, NPA752_139}. In the strangeness exchange reaction, the
strange quark is brought in by the kaon beam and subsequently
transferred to the target to form a hypernucleus. By choosing the
incident momentum of the kaon, the momentum transfer can be very
small for small $\pi^-$ angles in this reaction, which favors the
formation of low spin states of the hypernucleus. The use of
energetic pion beams for the hypernucleus production was first
proposed by Dover, Ludeking and Walker\cite{PRC22_2073, Ludeking}
and further studied by Bando et al \cite{PTP76_1321}. The associated
$(\pi ^+,K^+)$ reaction, in which a pair of s and $\bar{s}$ quarks
are produced, is endoergic, so the momentum transfer should be
larger than the Fermi momentum for a pion momentum with which the
two-body $\pi ^+n \rightarrow K^+\Lambda$ cross section remains
sizeable. The experimental studies of the $(\pi ^+,K^+)$ reactions
at AGS and KEK clearly demonstrated the usefulness of this reaction
for investigating the excitation of hypernuclei\cite{EPLA18_283,
Tamura, PTPS156_104}.

 The hypernucleus can also be produced in p-A collisions. The
possibility to produce hypernuclei in the $(p,K^+)$ reaction was
firstly mentioned by Yamazaki\cite{yam}.
 The possible associated-production reactions for hypernuclei in p-A
 collisions include the following types\cite{198001280},
\begin{eqnarray}
1)~ ~p+_AZ &\rightarrow& _{\Lambda}^{A+1}Z+K^+ ,\nonumber\\
2)~ ~p+_AZ &\rightarrow& _{\Lambda}^A(Z-1)+p+K^+, \nonumber\\
3)~ ~p+_AZ &\rightarrow& K^+ +X.\nonumber\\
\end{eqnarray}
Reaction 1) is a process with a two-body final state and is  more
suitable for high-resolution studies.  Theoretical studies of the
$A(p,K^+){_\Lambda B}$ reaction were performed for some nuclei. The
exclusive $K^+$ meson production in a proton-nucleus collision,
leading to a two-body final state, is investigated in a fully
covariant two-nucleon model based on an effective Lagrangian
picture\cite{0505043}. However, due to the large momentum transfer
occurred in the reaction, the reaction cross sections are very
small, which makes the experimental study very difficult. Reaction
3) is an inclusive process, where the outgoing $K^+$ is measured
while the rest of the system in the final state X is unobserved.
Some experiments for proton-induced-hypernucleus productions have
been proposed in several laboratories\cite{PLB427_403, EPJA11_1,
APPB33_603, NPA585_91}, but the experimental data and corresponding
theoretical works are still sparse at this
moment\cite{NPA450_147c,ZPA351_411,JPG21_L69,9606059,NPA639_177c}.
The production of $\Lambda$ hypernuclei was studied in the reaction
of the proton with the nucleus, and the lifetimes of the produced
heavy hypernuclei were measured by the observation of delayed
fission using the recoil shadow
method\cite{PLB427_403,EPJA11_1,APPB33_603} at COSY. At the same
time, the experimental study of the production and decay of
hypernuclei in p-A collisions  was also carried out at
Dubna\cite{NPA585_91}.

Reaction 2) is a hypernuclear production process with a three-body
final state in the p-A collision. This reaction was suggested for
producing hypernuclei 35 years ago by Fetisov\cite{PLB38_129}. By
choosing the momenta of the outgoing proton and $K^+$-meson, one can
transfer a smaller momentum to the $\Lambda$ hyperon, which is
favorable for the formation of the $\Lambda$-hypernucleus.  In the
early work of Fetisov, the cross section of this reaction was
estimated in the framework of the plane-wave impulse approximation
and the cross section of the elementary process was taken from the
early experimental data. It was found that the cross sections of
these reactions are measurable with modern experimental technique.

In recent years, with the development of experimental technologies
and methods, the proton beam with a substantially larger intensity
is available. It provides an opportunity to re-examine the
hypernuclear production in the three-body-final-state p-A collision.
In this paper, we investigate this reaction in the framework of the
distorted-wave-impulse-approximation (DWIA). The elementary process
$p+p \rightarrow p+K^+ +\Lambda$ is calculated in a fully covariant
two-nucleon model, where the effective Lagrangian, including
contributions from the $N(1535)$, $N(1650)$, $N(1710)$ and $N(1720)$
excitations in the intermediate state, is employed.

In section 2, the theoretical model for the $A(p,pK^+){_\Lambda B}$
reaction is briefly illustrated. Calculated results and analysis for
the $A(p,pK^+){_\Lambda B}$ reaction are presented in section 3. In
section 4 the concluding remarks are given .

%%%%%%%%%%%%%%%%%%%%%%%%%%%%%%%%%%%%%%%%%%%%%%%%%%%%%%%%%%%%%%%%%%%%%%%%%%%
%section 2
\section{Theoretical model for $A(p,pK^+){_\Lambda B}$ reactions}

\begin{figure}[h]
\centering \includegraphics[]{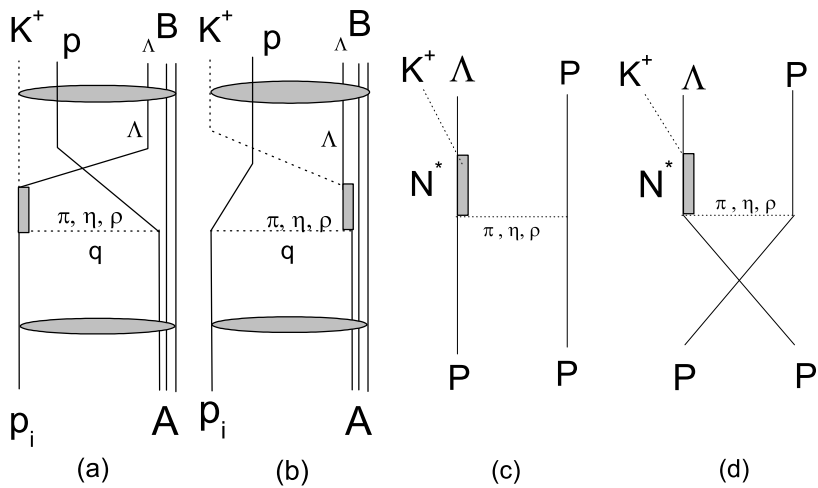} \caption{Ferynman
diagrams for $ p A \rightarrow p K^+  {_\Lambda B}$ and $p
p\rightarrow p K^+ \Lambda $ reactions. The elliptic shaded areas
represent the interactions in the incoming and outgoing channels.
The rectangular shaded areas stands for nucleon
resonances.}\label{feynman}
\end{figure}

The production reaction of the $\Lambda$-hypernucleus in the p-A
collision can proceed via a one-step process $pN \rightarrow K^+
\Lambda N$ or a two-step cascade process in which a pion or a
$\triangle$ is created in the intermediate stage, namely $pN
\rightarrow \pi NN$ and $\pi N \rightarrow K^+ \Lambda$ or $pN
\rightarrow \triangle N$ and $\triangle N \rightarrow K^+ \Lambda
N$. However, it is commonly believed that the formation of the
hypernucleus in the p-A collision is dominated by the one-step
mechanism. Based on the one-step mechanism $pp\rightarrow pK^+
\Lambda$, we study the $A(p,pK^+){_\Lambda B}$ reaction in the
framework of DWIA. In the reaction, we assume that the incoming
proton interacts with one of the protons in the target and a $N^*$
resonance is excited and followed by its decays into a $K^+$ meson
and a $\Lambda$ particle. Then, by the final-state interaction (FSI)
between the $\Lambda$ and the residual nucleus, the
$\Lambda$-hypernucleus is formed. The relevant Feynman diagrams for
the $A(p,pK^+){_\Lambda B}$ reaction are depicted in
Fig.\ref{feynman}, where Figs.\ref{feynman}(a) and (b) stand for the
direct process in which the incident proton is excited into a $N^*$
and for the exchange process which corresponds to the excitation of
the target proton, respectively.

In the framework of DWIA the transition matrix for the
$A(p,pK^+)_{\Lambda}B$ reaction can be written as
\begin{equation}
    T_{fi}^{pA\rightarrow pK^+{_{\Lambda} B}}= \langle \Phi_{f}(\vec{r}_{K},
    \vec{r}_0,\vec{r}_1,...,\vec{r}_A)|\widehat{O}(\vec{r}_{K},\vec{r}_0,\vec{r}_1,
    \vec{r}_2,...,\vec{r}_A)
    |\Phi_i(\vec{r}_0,\vec{r}_1,...,\vec{r}_A)\rangle
\end{equation}
with
\begin{equation}\label{5}
    \Phi_i(\vec{r}_0,\vec{r}_1,...,\vec{r}_A)=\chi_p^{(+)}
    (\vec{P}_p,\vec{r}_0)\Psi_i(\vec{r}_1,...,\vec{r}_A),
\end{equation}
\begin{equation}\label{6}
        \Phi_f(\vec{r}_0,\vec{r}_1,...,\vec{r}_A)=\chi_{p'}^{(-)}
        (\vec{P}_{p'},\vec{r}_0)\chi_{K}^{(-)}(\vec{P}_{K},\vec{r}_{K})
    \Psi_f(\vec{r}_1,...,\vec{r}_A),
\end{equation}
where $\Phi_i(\vec{r}_0,\vec{r}_1,...,\vec{r}_A)$ and
$\Phi_f(\vec{r}_{K},\vec{r}_0,\vec{r}_1,...,\vec{r}_A)$ are the wave
functions of the initial and final states, respectively.
$\Psi_i(\vec{r}_1,...,\vec{r}_A)$ and $
\Psi_f(\vec{r}_1,...,\vec{r}_A)$ denote the wave functions of the
target nucleus and the hypernucleus, respectively.
$\chi_p^{(+)}(\vec{P}_p,\vec{r}_0)$,
$\chi_{p'}^{(-)}(\vec{P}_{p'},\vec{r}_0)$ and
$\chi_{K}^{(-)}(\vec{P}_{K},\vec{r}_{K})$ represent the distorted
wave functions of the incoming proton, the outgoing proton and the
outgoing $K^+$ meson, respectively. $\vec{P}_p$ is the relative
momentum of the incoming proton in the proton-nucleus c.m. system.
$\vec{P}_{p'}$ and $\vec{P}_{K}$ are the momenta of the outgoing
proton and $K^+$ meson, respectively. The transition operator
$\widehat{O}$ can be written as,
\begin{equation}\label{3}
    \widehat{O}(\vec{r}_{K},\vec{r}_0,\vec{r}_1,\vec{r}_2,...,\vec{r}_A)=
    \sum_{i=1}^Z v_{0i}(\vec{r}_0,\vec{r}_i,\vec{r}_{K}),
\end{equation}
where $v_{0i}$ is the interaction for the production of a $K^+$
meson in the collision of the incoming proton with a proton inside
the nucleus, which is related to the t-matrix of the $pp\rightarrow
pK^+\Lambda$ reaction in the impulse approximation. Z is the number
of protons involved in the production process in the target nucleus.

Because the range of the nucleon-nucleon interaction is short
compared with the extension of the nuclear wave function, one may
assume that the incoming proton interacts with one of the nucleons
in the target while the rest of the nucleons remain unchanged, which
is a usual way to separate a single nucleon wave function from the
many-body system. Then, the transition T-matrix for the hypernucleus
production reaction reads as
\begin{equation}
T^{pA \rightarrow pK^+ {_\Lambda B}}=Z \cdot
F_{fi}(\vec{P}_{p'},\vec{P}_{K}; \vec{P}_p)\cdot t_{pp\rightarrow
p\Lambda K^+},
\end{equation}
where $t_{pp \rightarrow p\Lambda K^+}$ stands for the t-matrix of
the elementary $K^+$ production in p-p collisions and
\begin{equation}
F_{fi}(\vec{P}_{p'},\vec{P}_{K}; \vec{P}_p )=\int
dV\chi^{(-)*}_{p'}(\vec{P}_{p'},\vec{r})
\chi^{(-)*}_{K}(\vec{P}_{K},\vec{r})
\Phi^*_{J_{H}M_{H}T_{H}M_{T_{H}}}\Phi_{J_{A}M_{A}T_{A}M_{T_{A}}}
\chi^{(+)}_p(\vec{P}_p ,\vec{r})\label{formfactor}
\end{equation}
stands for the transition form factor that takes into account the
effects from the distorted waves of incoming and outgoing particles
and the wave functions of the target nucleus and hypernucleus. For a
simple target nucleus where $n$ nucleons stay in the $j$-subshell,
we write the wave function of the nucleus as
\begin{eqnarray}
\Phi_{J_{A},M_{A},T_{A},M_{T_{A}}}(j^n(\alpha))=\sum_{\alpha_0,J_0,T_0}
<j^{n-1}(\alpha_0,J_0,T_0)(l)j\frac{1}{2}J_AT_A\mid\}j^n(\alpha)J_A M_A T_A M_{T_A}>\nonumber\\
\times\sum_{M_0,M_{T_0}}C_{jM_A-M_0,~J_0~M_0}^{J_AM_A}C_{1/2M_{T_A}-M_{T_0},~T_0M_{T_0}}^{T_AM_{T_A}}
\phi_{jM_A-M_0,~1/2M_{T_A}-M_{T_0}}\Phi_{J_0M_0,~T_0M_{T_0}}(j^{n-1}(\alpha_0)),
\end{eqnarray}
where$<j^{n-1}(\alpha_0,J_0,T_0)(l)jJ_AT_A\mid\}j^n(\alpha)J_A M_A
T_A M_{T_A}>$ is the fractional parentage coefficient (f.p.c.). For
the hypernucleus, we assume a weak coupling scheme, namely the
$\Lambda$ particle couples with a nuclear core to form a
hypernuclear state, and express the wave function of the
hypernucleus as
\begin{eqnarray}
\Phi_{J_{H}M_{H}T_0M_{T_0}}(j^{n-1}(\alpha_0))=\sum_{M_0}C_{j_{\Lambda}M_H-M_0,~J_0M_0}^{J_HM_H}
\phi_{j_{\Lambda}M_H-M_0}\Phi_{J_0M_0,~T_0M_{T_0}}(j^{n-1}(\alpha_0)).
\end{eqnarray}
Subsequently, the transition form factor becomes
\begin{eqnarray}
&~&F_{fi}(\vec{P}_{p'},\vec{P}_{K}; \vec{P}_p)=
<j^{n-1}(\alpha_0,J_0,T_0)(l)jJ_AT_A\mid\}j^n(\alpha)J_AM_AT_AM_{T_A}> \nonumber\\
&\times& \sum_{M_0,M_{T_0}}C_{jM_A-M_0,~J_0~M_0}^{J_AM_A}
C_{1/2M_{T_A}-M_{T_0},~T_0M_{T_0}}^{T_AM_{T_A}}
C_{j_{\Lambda}M_H-M_0,~J_0M_0}^{J_HM_H} \nonumber\\
&\times& \int d^3r\chi^{(-)*}_{p'}(\vec{P}_{p'},\vec{r})
\chi^{(-)*}_{K}(\vec{P}_{K},\vec{r})
\phi^*_{j_{\Lambda}M_H-M_0}(\vec{r})\phi_{jM_A-M_0,~1/2M_{T_A}-M_{T_0}}(\vec{r})
\chi^{(+)}_p(\vec{P}_p ,\vec{r})\label{formfactor}.
\end{eqnarray}
Moreover, as a very good approximation at higher
energies\cite{PRC47_1701, PLB378_29}, we adopt the eikonal
approximation to describe the distorted waves:
%It is well known that for the distorted wave function, the eikonal
%form is a very good approximation at the higher energy\cite{oset}.
%We therefore adopt the eikonal approximation to describe the
%distorted waves:
\begin{eqnarray}
&\chi^{(+)}_p(\vec{P}_{p},\vec{r})=e^{i\vec{P}_p\cdot\vec{r}}
\exp\left[-\frac{\sigma_p}{2}\int^z_{-\infty}\rho(b,z')dz'\right]\label{wf3},\\
&\chi^{(-)*}_{p'}(\vec{P}_{p'},\vec{r})=e^{-i\vec{P}_{p'}\cdot\vec{r}}
\exp\left[-\frac{\sigma_p}{2}
\int_0^{\infty} \rho(b,z+l\frac{\vec{P}_{p'}}{P_{p'}})dl\right]\label{wf1}, \\
&\chi^{(-)*}_{K}(\vec{P}_{K},\vec{r})=e^{-i\vec{P}_{K}\cdot\vec{r}}
\exp\left[-\frac{\sigma_K}{2}\int_0^{\infty}
\rho(b,z+l\frac{\vec{P}_{K}}{P_{K}})dl\right]\label{wf2},
\end{eqnarray}
where $\sigma_p$ and $\sigma_K$ are the $p-N$ and $K^+-N$ total
cross sections, respectively, and $\rho$ the density distribution of
the target nucleus.

The differential cross section of the $A(p,pK^+)_{\Lambda}B$
reaction with the initial nuclear state $J_A$ and the final
hypernucleus state $J_H$ can be written as
\begin{equation}
\begin{aligned}
d\sigma^{((l)jJ_A,J_H)}(pA \rightarrow pK^+ {_\Lambda
B})&=\frac{(2\pi)^4}{4\sqrt{(P_p\cdot
P_A)^2-(m_pm_A)^2}}\overline{\sum_{s_i}}\sum_{s_f}|T_{J_H,J_A}^{pA
\rightarrow pK^+ {_\Lambda B}}|^2&d\phi_3 , \label{eq:difcs}
\end{aligned}
\end{equation}
where $d\phi_3$ is the differential three-body phase space, $S_i$
and $S_f$ denote the spins of the initial and final states,
respectively, $P_p$ and $P_A$ are the four momenta of the incoming
proton and the target nucleus, respectively, and $m_p$ and $m_A$ are
the masses of the proton and the nucleus, respectively. By
integrating over relevant variables in Eq.(\ref{eq:difcs}), the
measurable physical quantities, such as differential cross sections
for the outgoing proton, kaon and hypernucleus, double differential
cross sections, missing mass spectra of the reaction and etc., can
be formulated. For example, the differential cross section and
double differential cross section for the outgoing kaon can be
written as
\begin{align}
\frac{d\sigma}{d\Omega_{K}}=\int\frac{E_{p}\text{\/}E_{A}\text{\/}E_{K}%
E_{p'}\text{\/}E_{H}}{(2\pi)^{5}|p_{pcm}|\text{\/}s}\overline{\sum_{s_i}}\sum_{s_f}|T_{J_H,J_A}^{pA
\rightarrow pK^+ {_\Lambda
B}}|^2|\vec{P}^{\ast}_{p'}||\vec{P}_{K}|dm_{pH}d\Omega^{\ast}_{p'},
\label{eq:difcsk}
\end{align}

\begin{align}
\frac{d\sigma}{dP_{K}d\Omega_{K}}=\int\frac{E_{p}\text{\/}E_{A}\text{\/}|\vec{P}_{K}|^{2}|\vec{P}_{p'}|^{2}}
{(2\pi)^{5}|p_{pcm}|\sqrt{s}}\frac{1}{|\frac{dG}{d|\vec{P}_{p'}|}|}\overline{\sum_{s_i}}\sum_{s_f}|T_{J_H,J_A}^{pA
\rightarrow pK^+ {_\Lambda B}}|^2d\Omega_{p'}, \label{eq:difcskk}
\end{align}
where $m_{pH}$ is the invariant mass of the outgoing proton and
hypernucleus; $p_{pcm}$ and $\sqrt{s}$ are the incident c.m.
momentum and total energy; $\frac{dG}{d|\vec{P}_{p'}|}$ can be
written as
\begin{align}
\frac{dG}{d|\vec{P}_{p'}|}=-\frac{|\vec{P}_{p'}|}{E_{p'}}-\frac{|\vec{P}_{p'}|+|\vec{P}_{K}|(\sin
\theta_{p'} \cdot \sin\theta_{K} \cdot \cos
(\varphi_{p'}-\varphi_{K})+\cos \theta_{p'} \cdot
\cos\theta_{K})}{E_{H}}.
\end{align}
Similar to the above derivations, we can obtain the single and
double differential cross sections for the outgoing proton and
hypernucleus, respectively. By integrating over all possible solid
angles, we can further get double differential cross sections
\begin{align}
\frac{d\sigma}{dP_{p}dP_{K}}=\int\frac{E_{p}\text{\/}E_{A}\text{\/}%
E_{H}|\vec{P}_{p'}||\vec{P}_{K}|}{(2\pi)^{5}|P_{pcm}|\sqrt{s}}\overline{\sum_{s_i}}\sum_{s_f}|T_{J_H,J_A}^{pA
\rightarrow pK^+ {_\Lambda B}} |^{2} d\Omega_{p'}d\varphi_{p'K},
\label{eq:doubpp}
\end{align}
where the $\varphi_{p'K}$ is the polar direction of $\vec{P}_K$ with
respect to $\vec{P}_{p'}$. On the other hand, if we define invariant
missing mass as $m_x^2=(P_0+P_A-P_p-P_K)^2$, the missing mass
spectra for the four-body-final-state reaction of $p+A\rightarrow p+
K^+ +\Lambda+ B$ will be given as,
\begin{align}
\frac{d\sigma}{dm_{x}}=\int\frac{\pi^{2}E_{p}\text{\/}E_{A}\text{\/}%
E_{p'}E_{K}E_{\Lambda}\text{\/}E_{A'}}{16(2\pi)^{8}|p_{pcm}|s^{2}p_{p'K}}\overline{\sum_{s_i}}\sum_{s_f}|T_{J_H,J_A}^{pA
\rightarrow pK^+ {_\Lambda B}} |^{2} d\Omega_{p'}
d\varphi_{p'K}d\varphi_{\Lambda}dm_{\Lambda
A^{\prime}}^{2}dm_{p'KA^{\prime}}^{2}dm_{p'\Lambda A^{\prime}}^{2},
\end{align}
where $p_{p'K}=|\vec{P}_{p'}+\vec{P}_{K}|$; $\varphi_{\Lambda}$ is
the polar direction of $\vec{P}_{\Lambda}$ with respect to
$\vec{P}_{p'}+\vec{P}_{K}$; $m_{\Lambda A^{\prime}}$,
$m_{p'KA^{\prime}}$ and $m_{p'\Lambda A^{\prime}}$ are respectively
invariant masses and defined as $m^2_{\Lambda
A^{\prime}}=(P_{\Lambda}+P_{A'})^2$,
$m^2_{p'KA^{\prime}}=(P_{p'}+P_K+P_{A'})^2$ and $m^2_{p'\Lambda
A^{\prime}}=(P_{p'}+P_{\Lambda}+P_{A'})^2$.

Finally, by summing up all the quantum numbers for considered
initial states of the proton in the target and possible states of
the hypernucleus, the total reaction cross sections can be obtained
by
\begin{equation}
\sigma^{tot}(p+A \rightarrow p+K^+ +{_\Lambda B})=\sum_{(l)j,J_H}
\sigma^{((l)jJ_A,J_H)}(p+A \rightarrow p+K^+ +{_\Lambda B}).
\end{equation}

%%%%%%%%%%%%%%%%%%%%%%%%%%%%%%%%%%%%%%%%%%%%%%%%%%%%%%%%%%%%%%%%%%%%%%%%%%%
%section 3
\section{Numerical Results and Discussions }

\begin{figure}[h]
\centering \includegraphics[]{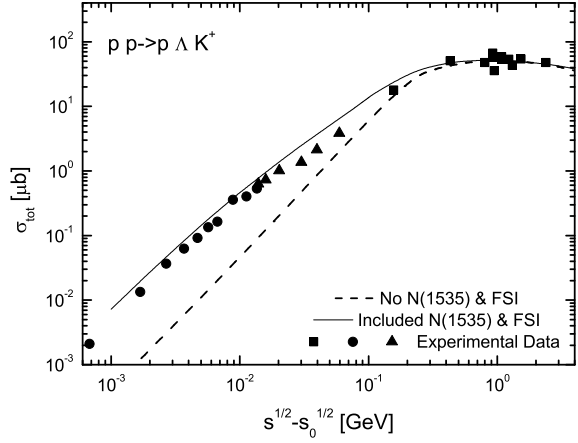} \caption{Total cross section
of the $pp \rightarrow p K^+ \Lambda$ reaction. $s^{1/2}$ and
$s^{1/2}_0 = m_N + m_{\Lambda} + m_K$ are the invariant collision
energy and threshold energy, respectively. The experimental data are
taken from Ref.\cite{Landolt}. The solid and dashed curves denote
the results with and without the contributions of $N^*(1535)$ and
$p\Lambda$ FSI, respectively.}\label{ppkl}
\end{figure}

We first discuss the t-matrix of the elementary process
$pp\rightarrow pK^+ \Lambda$ which has been studied by many
authors\cite{9801063, PRL96_042002, 0004022, PRC60_055213} within
the framework of the meson exchange theory. For example, Tsushima
\textit{et al.}\cite{9801063} studied such a process by including
the contributions from the $N^*(1650)$, $N^*(1710)$ and $N^*(1720)$
in the resonant model. The resultant total cross sections well agree
with the experimental data at high energies. In order to better
describe the data in the threshold region, an additional
sub-threshold $N^*(1535)$ resonance was suggested by Liu \textit{et
al.}\cite{PRL96_042002}. It was found that the
$N^*(1535)-\Lambda-K^+$ coupling is quite large and the contribution
from $N^*(1535)$ is very important for the cross sections of the $pp
\rightarrow p K^+ \Lambda$ reaction near the threshold. Based on
these models, we re-calculate the amplitude of this elementary
process by further considering the $p-\Lambda$ final-state
interaction (FSI) \cite{0004022}. In the calculation, the effective
Lagrangians are chosen as those in Refs.\cite{PRL96_042002,9801063}.
The resultant total cross sections for the $pp \rightarrow p K^+
\Lambda$ reaction are plotted together with the experimental data
\cite{Landolt} in Fig.\ref{ppkl}, where the solid and dashed curves
denote the results with and without the contributions of the
$N^*(1535)$ and the $p-\Lambda$ FSI, respectively. It is shown that
by taking the contribution from the $N^*(1535)$ resonance and the
$p-\Lambda$ FSI into account, the experimental data can be well
described in both the high energy and low energy regions.

What we would like to know in this paper is that if concerned
hypernuclei do exist, can we observe them in the $p+A \rightarrow
p+K^+ +_\Lambda B$ reaction and accurately measure their mass
spectrum for the study of the hyperon-nucleon (Y-N) interaction in
future. For this purpose, we first employ a phenomenological
$\Lambda$-nucleus potential to fit the masses of observed
hypernuclei and obtain the bound state wave functions of the
hypernuclei which will be studied in this paper. And then, with
these wave functions, we calculate the measurable physical
quantities to see if we can achieve our expectations.

As discussed in last section, we assume that the $\Lambda$ particle
interacts with a nucleus core to form a hypernuclear state . The
bound-state wave function of the $\Lambda$ is obtained by solving
the Schr\"{o}dinger equation with a phenomenological
$\Lambda$-nucleus potential\cite{PRC38_2700}
\begin{equation}
 U_{\Lambda -A}(r)=-U_0\frac{1}{1+e^{(r-R_0)/a}},
\end{equation}
where $U_0$=28MeV, $R_0=(1.128+0.439A^{-2/3})$fm and a=0.54fm. The
resultant binding energies of hypernuclei as well as the
experimental data are tabulated in the second column in Table
\ref{BEnergy}.

%%%%%%%%%%%%%%%%%%%%%%%%%%%%%%%%%%%%%%%%%%%%%%%%%%
\begin{table}[h]
\caption{\label{BEnergy} The binding energies of
$\Lambda$-hypernuclei. The corresponding experimental data are
taken from\cite{PRC38_2700,NPA585_211c}. The $l_{\Lambda}$-state
denoted the orbital angular momentum states of $\Lambda$-hyperon
relative to core nucleus.}
\begin{center}
\begin{tabular}{|c|c|c|}
\hline
Hypernucleus&\multicolumn{2}{|c|}{Binding Energy(MeV)}\\
\cline{2-3} ($l_{\Lambda}$-state)&solution of
  & Exp.\\
 & Schr\"{o}dinger equation  & \\\hline
$^6_{\Lambda}He$($s_{\Lambda}$-state)  & 6.06      &4.18\\
\hline
$^{12}_{\Lambda}B$($s_{\Lambda}$-state) &11.03     &11.37 \\
~~~~($p_{\Lambda}$-state)  & 0.48    &\\
\hline
$^{16}_{\Lambda}N$($s_{\Lambda}$-state)  & 12.97      & (13)\\
~~~~($p_{\Lambda}$-state)  & 2.51     &\\
\hline
\end{tabular}
\end{center}
\end{table}
%%%%%%%%%%%%%%%%%%%%%%%%%%%%%%%%%

Then, we calculate the reaction cross sections of a proton
bombarding on $^6Li$,$^{12}C$ and $^{16}O$ targets. Same as that
used in Refs.\cite{NPA368_503,CPL18_1030}, for the light target
nucleus, we use the simple single-particle wave function of harmonic
oscillator, $\phi_{l_{1}m_{1}}$, to describe target nucleons. For
simplicity, we approximately take
$<j^{n-1}(\alpha_0,J_0,T_0)(l)jJ_AT_A\mid\}j^n(\alpha)J_AT_A>=1$.

The nuclear distortion effects in the initial and final states are
considered by using the eikonal approximation. We use the form of
the harmonic-oscillator density(HO) for the nuclear charge
distribution,
\begin{eqnarray}
\rho(r)=\rho_0~[1+\alpha (\frac{r}{a})^2]exp[-(\frac{r}{a})^2]
\end{eqnarray}
with the parameters determined from electron
scatterings\cite{ADANDT14_479, ADNDT36_495}. Again for simplicity,
we assume that the distributions for protons and neutrons in nuclei
are the same. The total cross sections of two-body interactions,
$\sigma_{p}$ and $\sigma_{K}$, are taken from the experimental data
at the corresponding relative momenta of the incident proton and
outgoing kaon with respect to the target and residual nucleus,
respectively\cite{Landolt}.

It is worthy to mention that the differential reaction cross section
is very sensitive to the form factor, and consequently to the
momentum transfer, because the form factor is a function of the
momentum transfer. When the momentum transfer increases, the form
factor decreases rapidly. Thus, in order to reach a relatively large
cross section for the hypernucleus production in the experiment, the
energy of the incident proton should be limited in a range where the
momentum transfer has the smallest value.
\begin{figure}[h]
\centering \includegraphics[]{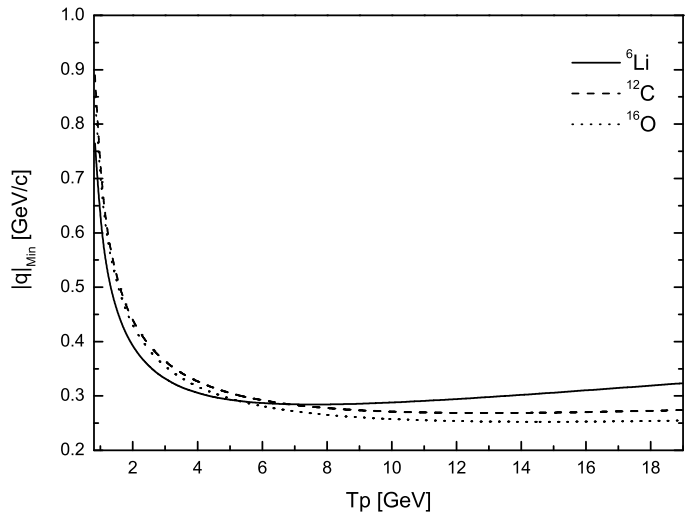} \caption{The minimum
momentum transfers for reaction $^{6}Li(p,pK^+){^{6}_{\Lambda}
He}$, $^{12}C(p,pK^+){^{12}_{\Lambda} B}$ and
$^{16}O(p,pK^+){^{16}_{\Lambda} N}$.}\label{momentumq}
\end{figure}
In Fig.\ref{momentumq}, we show the minimum momentum transfer as a
function of the kinetic energy of the incident proton for the
reactions $^{6}Li(p,pK^+){^{6}_{\Lambda} He}$,
$^{12}C(p,pK^+){^{12}_{\Lambda} B}$ and
$^{16}O(p,pK^+){^{16}_{\Lambda} N}$, respectively. We find that the
minimum momentum transfer decreases with the increasing energy of
the incident proton from the threshold of the reaction. When the
kinetic energy of the proton is larger than 2 GeV, which is the
available energy of the proton beam at COSY and the Cooling Storage
Ring (CSR) at Lanzhou in China, the minimum momentum transfer
becomes smaller than $300-400 $MeV/c.

\begin{figure}[h]
\centering \includegraphics[]{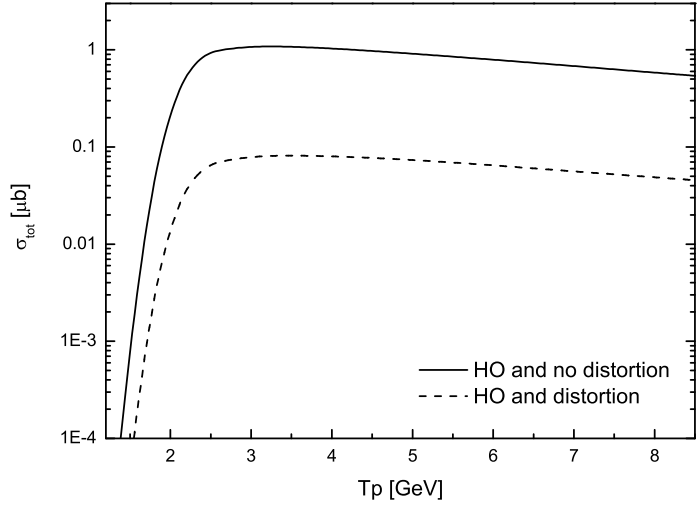} \caption{The total
cross section of the $^{12}C(p, pK^+)^{12}_{\Lambda}B$ reaction as a
function of the incident proton energy. The solid and dashed curves
represent the calculated results of TCS within the frameworks of
PWIA and DWIA, respectively.}\label{TCSC12DandND}
\end{figure}

The calculated total cross sections(TCS) of the
$^{12}C(p,pK^+){^{12}_{\Lambda} B}$ reaction in the framework of
DWIA are presented in Fig.\ref{TCSC12DandND} and compared with the
results in PWIA. In the calculation, the nuclear density in the HO
form is employed for the evaluation of the distortions. We find that
the distortion effects tend to reduce the cross section
significantly by a factor of 3 to 10, and this effect will be more
pronounced as the target nucleus becomes heavier.

In fact, we are not able to detect the hypernucleus directly in
experiment. The commonly used method to confirm the production of
the hypernucleus is to detect the other concomitant particles. In
the $A(p,pK^+){_\Lambda B}$ reaction, it is feasible to detect the
outgoing proton and kaon.

\begin{figure}[h]
\centering \includegraphics[]{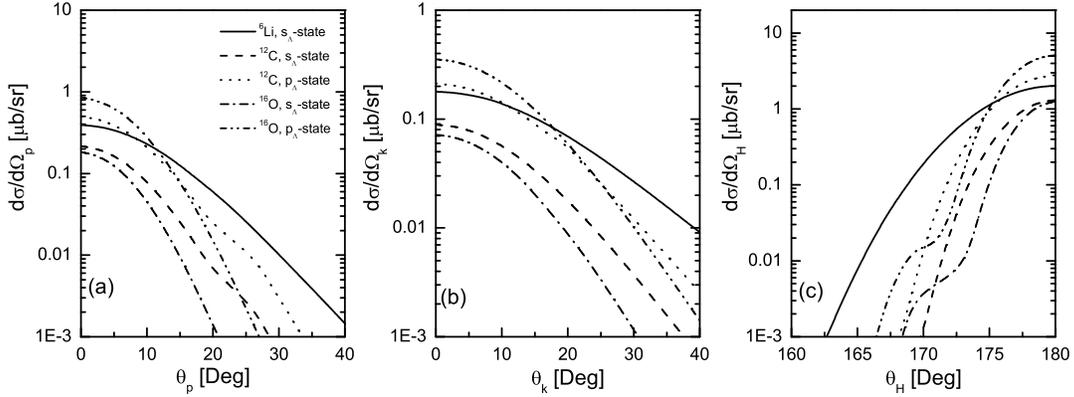} \caption{The differential
cross sections in the c.m. system for the $^6Li$, $^{12}C$ and
$^{16}O$ target nuclei in the framework of DWIA. The kinetic energy
of the incident proton, $T_p$, is 2.7GeV. }\label{DCSAll}
\end{figure}

The differential cross sections (DCS) in the c.m. system for the
$^6Li$, $^{12}C$ and $^{16}O$ target nuclei in DWIA are presented in
Fig.\ref{DCSAll}, where diagrams (a), (b) and (c) denote the angular
distributions for the outgoing proton, kaon and hypernucleus,
respectively, and the kinetic energy of the incident proton, $T_p$,
is 2.7GeV.
From this figure, we find that the DCS for the outgoing proton, as
well as kaon, is larger at the forward angles, and for the resultant
hypernucleus it is larger at the backward angles, because of the
momentum conservation. For the same target, the DCS for the state
with the $p_{\Lambda}$-shell $\Lambda$ is generally larger than that
with the $s_{\Lambda}$-shell $\Lambda$. This can be understood in
the following way: Because the momentum transfer for the reaction
$A(p,pK^+){_{\Lambda} B}$ is still relatively large at $T_p$=2.7GeV,
the produced $\Lambda$ would obtain a larger momentum, and
consequently a higher orbital angular
momentum\cite{reaction2,PRC22_2037}.
 Then, the contribution from the
conversions of the $p$-shell and $s$-shell nucleons to a
$p_{\Lambda}$-shell $\Lambda$ is larger than that from the
conversion of the $p$-shell and $s$-shell nucleons to a
$s_{\Lambda}$-shell $\Lambda$. In the cases of $^6Li$, $^{12}C$ and
$^{16}O$ targets, the DCS of producing a $s_{\Lambda}$-state
hypernucleus is normally smaller for the heavier target (larger A)
than the one for the lighter target. This is because that the
magnitude of the wave function product $\Phi^*_{\Lambda}\Phi_{p}$
becomes smaller with increasing nucleon number $A$. Thus, in the
heavier target case, this factor, together with the distortion
effect which is relatively larger in the heavier target case, would
exceed the effect caused by the increased number of $p$-shell proton
and make the DCS reduced. Moreover, for different targets, the DCS
difference between the $p_{\Lambda}$-shell hypernucleus production
case and the $s_{\Lambda}$-shell hypernucleus production case for
the $^{16}O$ target is larger than that for the $^{12}C$ target. The
reason for this phenomenon is that in the case of the
$s_{\Lambda}$-shell hypernucleus production, as mentioned above, the
DCS in the $^{16}O(p$, $pK^+){^{16}_{\Lambda} N}$ reaction is
suppressed in comparison the one in the $^{12}C(p$,
$pK^+){^{12}_{\Lambda} B}$ reaction. However, in the case of the
$p_{\Lambda}$-shell hypernucleus production, the magnitude of
$\Phi^*_{\Lambda}\Phi_{p}$ for the $^{16}O(p$,
$pK^+){^{16}_{\Lambda} N}$ reaction is almost the same as that for
the $^{12}C(p$, $pK^+){^{12}_{\Lambda} B}$ reaction, and the number
of the $p$-shell proton in $^{16}O$ is larger than that in $^{12}C$,
so that the DCS for the $p_{\Lambda}$-shell hypernucleus production
in the $^{16}O(p$, $pK^+){^{16}_{\Lambda} N}$ reaction is larger
than that in the $^{12}C(p$, $pK^+){^{12}_{\Lambda} B}$ reaction.

\begin{figure}[h]
\centering \includegraphics[]{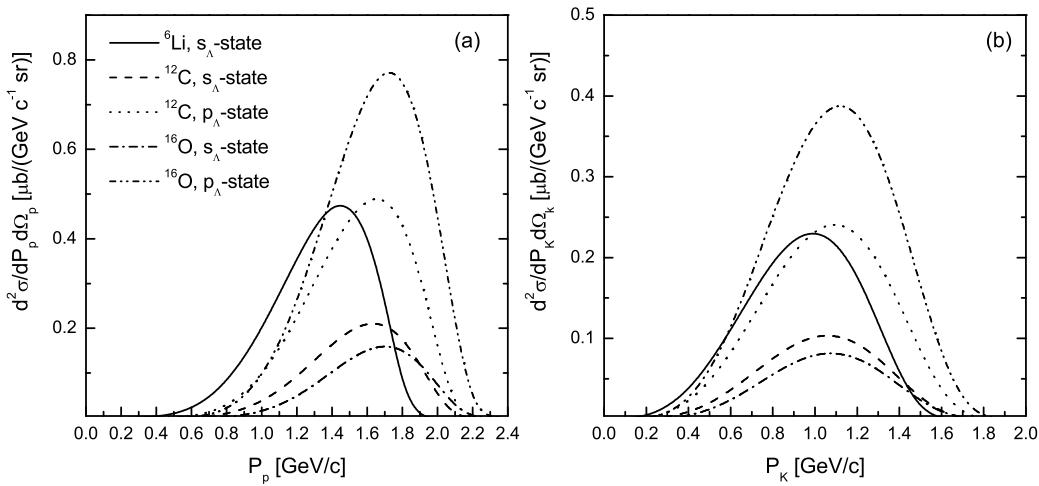} \caption{The double
differential cross sections versus (a) the momentum of the outgoing
proton at the angle $6^{\circ}$ and (b) the momentum of the outgoing
kaon at the angle $6^{\circ}$, in the C.M. system. The kinetic
energy of incident proton is 2.7GeV.}\label{DDCSAll2D}
\end{figure}

\begin{figure}[h]
\centering \includegraphics[]{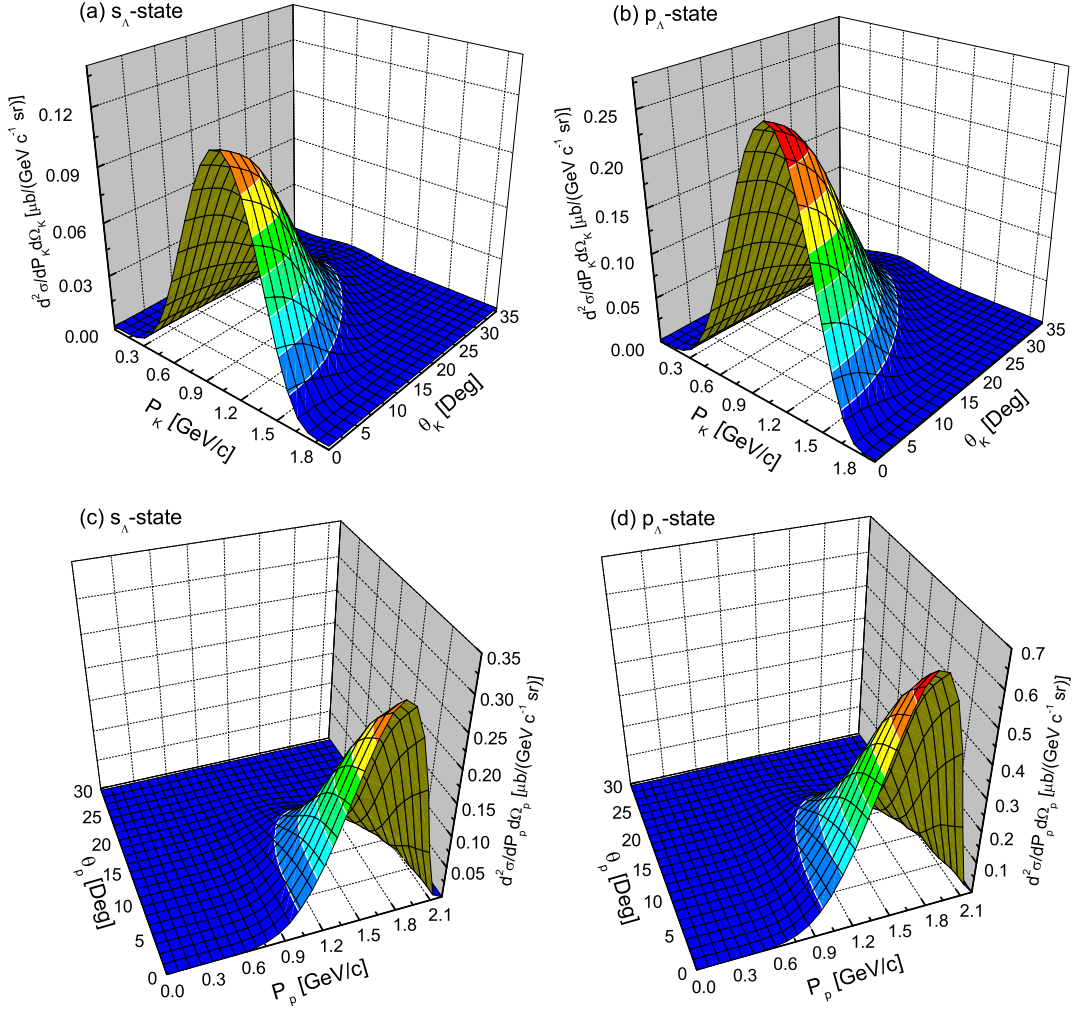} \caption{The 3D-plot
for DDCS with respect to the momentum and angle of outgoing kaon
((a) and (b)), and proton ((c) and (d)), where (a) and (c) for
$s_{\Lambda}$-shell hypernucleus and (b) and (d) for
$p_{\Lambda}$-shell hypernucleus.}\label{DDCSAll3D}
\end{figure}

In Fig.\ref{DDCSAll2D}, we show the dependance of the double
differential cross sections (DDCS) on the outgoing proton momentum
(a) and the outgoing kaon momentum (b), respectively.
The results show that for each hypernuclear production case, the
DDCS has a Gaussian-like distribution with respect to the momentum
of the outgoing proton or kaon. This means that there exist a
momentum range of outgoing proton and a momentum range of kaon,
where we can find the concomitant particles proton and kaon.
Furthermore, when the target becomes heavier, the maximum DDCS point
moves slightly to the higher proton momentum and kaon momentum
regions. This enable produced $\Lambda$ has smaller momentum so that
it can be bound on the core nucleus which has less recoil due to its
heavier mass. For a clear vision, we also present the DDCS with
respect to the momentum and angle of outgoing kaon, respectively,
for the $^{12}C$ target in Fig.\ref{DDCSAll3D}.

\begin{figure}[h]
\centering \includegraphics[]{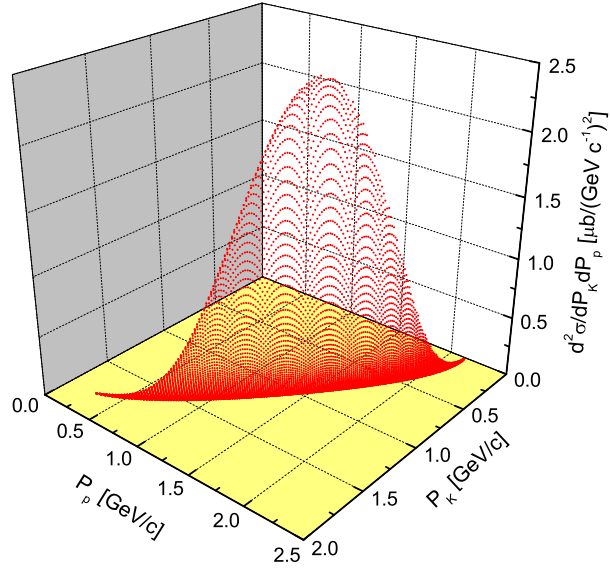} \caption{The c.m. double
differential cross section distribution with respect to the momenta
of the outgoing proton and kaon for the $^{12}C$ target nucleus in
DWIA. The kinetic energy of incident proton is
Tp=2.7GeV.}\label{DDCSpp}
\end{figure}

Since the final state of the reaction is a three-body state, one can
confirm the production of the hypernucleus by detecting the outgoing
proton and kaon simultaneously. We demonstrate the DDCS distribution
with respect to the momenta of the outgoing proton and kaon in
Fig.\ref{DDCSpp}
From this figure, we find that DDCS is distributed in a special
range due to the three-body phase space factor, which would be a
reference for the experiment.

\begin{figure}[h]
\centering \includegraphics[]{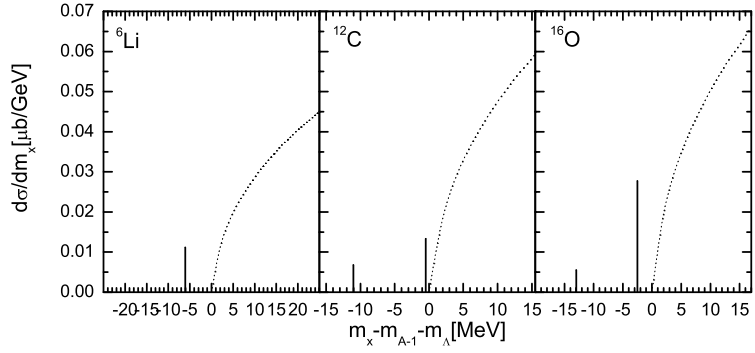} \caption{The missing mass
spectra for the $^{6}Li(p,pK^+){^{6}_{\Lambda} He}$,
$^{12}C(p,pK^+){^{12}_{\Lambda} B}$ and
$^{16}O(p,pK^+){^{16}_{\Lambda} N}$ reactions in DWIA at the kinetic
energy of the incident proton Tp=2.7GeV. The angles of outgoing
proton and kaon are limited in the range of $[0^{\circ},
12^{\circ}]$. }\label{mxAll}
\end{figure}

Furthermore, we study the mass and the formation signal of
hypernucleus by analyzing the missing mass spectrum in the inclusive
reaction $p+A\rightarrow p+ K^+  + X$. The invariant missing mass
can be defined as $m_x^2=(P_0+P_A-P-P_K)^2$. When the produced
$\Lambda$ can be bound on the core nucleus, namely the reaction
$A(p,pK^+){_{\Lambda} B}$ happens, the discrete peaks will appear in
the missing mass spectrum. The invariant missing mass $m_x$
corresponding to the peak position will associate with the mass of
the produced hypernuclear state. We calculate the missing mass
spectrum with the angles of outgoing proton and kaon to be limited
in the range of $[0^{\circ}, 12^{\circ}]$ and plot the missing mass
spectra for the $^{6}Li(p,pK^+){^{6}_{\Lambda} He}$,
$^{12}C(p,pK^+){^{12}_{\Lambda} B}$ and
$^{16}O(p,pK^+){^{16}_{\Lambda} N}$ reactions in the left, middle
and right diagrams of Fig.\ref{mxAll}, respectively. In the abscissa
of this figure, $m_{A-1}=m_A-m_p-E_{B_p}$, where $m_A$, $m_p$ and
$E_{B_p}$ are the mass of the target nucleus, the mass of the proton
and the binding energy of the last proton in the target,
respectively. One can see that each of the missing mass spectra
contain two parts: discrete lines and a continual curve. The
discrete lines express that the $\Lambda$ particle is in a bound
state with the residual nucleus, namely the final state is a
three-body state. The continual curve is due to the contribution
from a four-body final state. In Fig.\ref{mxAll}, the dotted curve
denotes four-body differential cross sections in each case. In the
left diagram, the single discrete line at the left-handed side of
the four-body DCS curve indicates that only one $s_{\Lambda}$-wave
hypernucleus state can be produced in the reaction with the $^6Li$
target. In the middle and right diagrams, the double discrete lines
at the left-handed side of the four-body DCS curve indicate that one
$s_{\Lambda}$-wave and one $p_{\Lambda}$-wave hypernuclear states
can be produced in the reaction with the $^{12}C$ and the $^{16}O$
targets, respectively.

In the calculation of the missing mass spectra we use the calculated
binding energies of the $\Lambda$ tabulated in Table\ref{BEnergy} to
obtain the masses of the hypernuclei. On the other hand, from the
measurement one can extract the masses of the produced hypernuclei.
 Therefore, if the
missing mass spectrum can accurately be measured, not only the
masses of the produced hypernuclear states can accurately be
determined, but also the hypernuclear production mechanism and the
detailed hyperon-nucleon interaction, and consequently the
hyperon-nucleus interaction, can authentically be studied.

%%%%%%%%%%%%%%%%%%%%%%%%%%%%%%%%%%%%%%%%%%%%%%%%%%%%%%%%%%%%%%%%%%%%%%%%
%section 4
\clearpage
\section{Conclusions}
We study the $A(p,pK^+)_{\Lambda}B$ reactions with several target
nuclei in PWIA and DWIA. The elementary process $pp\rightarrow pK^+
\Lambda$ is calculated in the resonance model in which the
contributions from the exchanges of the $\pi$, $\eta$ and $\rho$
mesons, as well as those from the intermediate resonant states
$N(1535)$, $N(1650)$, $N(1710)$ and $N(1720)$, are considered.
Especially, by additionally including the contribution from the
$N(1535)$ state and the $p\Lambda$ FSI, which provide a sizable
effect, the total cross section data of the elementary process in
the lower collision energy region can be much better reproduced.

In studying the $A(p,pK^+)_{\Lambda}B$ reactions in DWIA, the
distortion effects for incident proton and outgoing kaon are
obtained by the eikonal approximation. The $\Lambda$-nucleus wave
function is calculated by solving the Schr\"{o}dinger equation with
a Woods-Saxon potential, with which the empirical binding energies
of hypernuclei can be well reproduced.

The reaction cross sections for the $^6Li$, $^{12}C$ and $^{16}O$
targets are calculated. It is shown that the nuclear effects are
important and can reduce the cross section by factor of 3 $\sim$ 10.
The partial wave differential cross sections are nuclear mass
dependent. For the $s_{\Lambda}-$wave hypernucleus production, the
DCS is decreased with increasing nuclear mass, and the DCS for the
$p_{\Lambda}-$wave hypernucleus production is normally higher than
that for the $s_{\Lambda}-$wave hypernucleus production. However,
for the $p_{\Lambda}-$wave hypernucleus production, the DCS might
not be decreased with increasing nuclear mass, due to complicated
factors, such as the number of the $s_{\Lambda}-$shell protons in
the target, the magnitude of the product $\Phi^*_{\Lambda}\Phi_{p}$,
and etc.. The DDCS with respect to the momenta and the angles of the
outgoing proton and kaon are presented, respectively. The
three-dimensional plot for the DDCS with respect to the momenta of
outgoing proton and kaon is also demonstrated as a experimental
reference. It is noteworthy that the DDCS is distributed in a
special range due to the three-body phase space factor.

Finally, we propose an alternative way to measure the masses of
produced hypernuclei. That is the method of the missing mass
spectrum. We present the missing mass spectrum of the inclusive
process $p+A\rightarrow p+ K^+  + X$. From this spectrum, we can
accurately extract the masses of the hypernuclear states, if they
can be formed. As a consequence, we can authentically study the
hypernuclear production mechanism and the detailed hyperon-nucleon
interaction, and consequently the hyperon-nucleus interaction.

As an conclusion, we believe that the reaction
$A(p,pK^+)_{\Lambda}B$ is suitable for the hypernuclear physics
study. These reactions can be carried out in COSY and CSR due to the
$\mu$b-order reaction cross sections.

%%%%%%%%%%%%%%%%%%%%%%%%%%%%%%%%%%%%%%%%%%%%%%%%%%%%%%%%%%%%%%%%%%%%%%%%
\section{Acknowledgments}
One of the authors, Hantao Jing, is grateful to Xianhui Zhong and
Zhuoquan Zeng for some valuable discussions. This work is partially
supported by the National Natural Science Foundation of China under
grant Nos. 10775147, 10435080 and CAS grant No. KJCX3-SYW-N2.

%%%%%%%%%%%%%%%%%%%%%%%%%%%%%%%%%%%%%%%%%%%%%%%%%%%%%%%%%%%%%%%%%%%%%%%

%%%%%%%%%%%%%%%%%%%%%%%%%%%%%%%%%%%%%%%%%%%%%%%%%%%%%%%%%%%%%%%%%%%%%%%%
%References

%%%%%%%%%%%%%%%%%%%%%%%%%%%%%%%%%%%%%%%%%%%%%%%%%%%%%%%%%%%%%%%%%%%
%picture

%%%%%%%%%%%%%%%%%%%%%%%%%%%%%%%%%%%%%%%%%%%%%%%%%%%%%%%%%%%%%%%%%
\end{document}